\documentclass{article}
\usepackage{enumerate}
\usepackage{graphics}
\usepackage{amsmath}
\begin{document}
\begin {center}
{\bf{Extended Reaction Rate Integral as Solutions of Some General Differential Equations}}\\
\vskip.3cm
{D.P. Joseph}\\
\vskip.2cm
{\it Centre for Mathematical Sciences Pala Campus\\ Arunapuram P.O, Palai, Kerala 686 574, India}\\
and\\
\vskip.2cm
{H.J. Haubold}\\
\vskip.2cm
{\it Office for Outer Space Affairs, United Nations,\\
Vienna International Centre,\\P.O. Box 500, A-1400 Vienna,
Austria}\\
and\\
Centre for Mathematical Sciences, Pala Campus, Pala-686 574, Kerala, India\\[0.3cm]

\end{center}

\noindent
{\bf Abstract.} Here an extended form of the reaction rate probability
integral, in the case of nonresonant thermonuclear reactions with
the depleted tail and the right tail cut off, is considered. The
reaction rate integral then can be looked upon as the inverse of the
convolution of the Mellin transforms of Tsallis type statistics of
nonextensive statistical mechanics and stretched exponential as well
as that of superstatistics and stretched exponentials. The
differential equations satisfied by the extended probability
integrals are derived. The idea used is a novel one of evaluating
the extended integrals in terms of some special functions and then
by invoking the differential equations satisfied by these special
functions. Some special cases of limiting situations are also discussed.\\


 \noindent


\vskip.2cm
\section{Introduction}

\noindent

Nuclear reactions govern major aspects of the chemical evolution of
the universe. A proper understanding of the nuclear reactions that
are going on in hot cosmic plasma, and those in the laboratories as
well, requires a sound theory of nuclear-reaction dynamics.The
reaction probability integral is the probability per unit time that
two particles, confined to a unit volume, will react with each other
(Haubold and Mathai, 1998). In the present article we will show
that one can obtain the reaction rate probability integral as a
solution of a certain differential equation, the technique used is a
novel one. First we evaluate the integral and represent it as a
special function (Mathai, 1993; Luke, 1969). Then we can invoke
the differential equation satisfied by this special function,
thereby establishing the differential equation for the reaction rate integral. A study of this
integral as a solution of a differential equation is made because
the behavior of physical systems is usually studied with the help
of differential equations and hence the differential equations
derived in the present paper may become useful one day.\\
Here we will consider the integrals of the following form:\\

\noindent
{\it Nonresonant case with depleted tail}: (for details of
the integrals see Mathai and Haubold, 1988)
\begin{equation}
I_{1}^{(\delta)}=\int_{0}^{\infty}x^{\alpha-1}e^{-ax^{\delta}-bx^{-\rho}}{\rm
d}x,~~a>0,~b>0,~\delta>0,~\rho>0.
\end{equation}
{\it Nonresonant case with depleted tail and high energy cut-off}:
\begin{equation}
I_2^{(\delta)}=\int_{0}^{d}x^{\alpha-1}e^{-ax^{\delta}-bx^{-\rho}}{\rm
d}x,~~a>0,~b>0,~\delta>0,~\rho>0,~d<\infty.
\end{equation}
Do these integrals, as functions of $b$, satisfy some differential
equations? No such differential equations can be easily seen from
the integrals. We will show that these integrals will satisfy
certain differential equations. The main results will be stated as
follows. We will establish the following theorems for
$I_{1}^{(\delta)}$ and extended $I_{1\beta}^{(\delta)}$ and
$I_{2\beta}^{(\delta)}$. \\

\noindent
 {\bf\Large{Theorem 1}}\\

A constant multiple of the reaction rate integral in (1), under the
condition $\frac{\delta}{\rho}=m,m=1,2,\cdots$, can be obtained as a
solution of the differential equation,
\begin{equation}
\bigl[{(-1)}^{m+1}z-\eta\bigl(\eta-\frac{1}{m}\bigr)\cdots\bigl(\eta-\frac{m-1}
{m}\bigr)\bigl
(\eta-\frac{\alpha}{m\rho}\bigr)\bigr]f(z)=0,
\end{equation}
where,~$z=\frac{ab^m}{m^m},~\eta=z\frac{{\rm d}}{{\rm
dz}},~f(z)=\frac{\rho
a^{\frac{\alpha}{m\rho}}m^{\frac{1}{2}}}{(2\pi)^{\frac{1-m}{2}}}I_{1}^{(m\rho)}(z)$
 and
\begin{equation}
I_{1}^{(m\rho)}(z) =\frac{(2\pi)^{\frac{1-m}{2}}}{\rho
a^{\frac{\alpha}{m \rho}}m^{\frac{1}{2}}}
G^{m+1,0}_{0,m+1}\biggl[\frac{ab^m}{m^m}\bigg|_{0,\frac{1}{m},\cdots,\frac{m-1}{m},
\frac{\alpha}{m \rho}}\biggr],~~a>0,~b>0,~\alpha>0,~\rho>0.
\end{equation}
 \noindent
{\bf\Large{Corollary 1}}\\

When $\delta=1$, a constant multiple of the reaction rate integral
in (1), under the condition $\frac{1}{\rho}=m,m=1,2,\cdots$, can be
obtained as a solution of the differential equation,
\begin{equation}
\bigl[{(-1)}^{m+1}z-\eta\bigl(\eta-\frac{1}{m}\bigr)\cdots\bigl(\eta-\frac{m-1}{m}
\bigr)\bigl
(\eta-\alpha\bigr)\bigr]f(z)=0,
\end{equation}
where,~$z=\frac{ab^m}{m^m},~\eta=z\frac{{\rm d}}{{\rm dz}},
~f(z)=\frac{\rho
a^{\alpha}m^{\frac{1}{2}}}{(2\pi)^{\frac{1-m}{2}}}I_{1}(z)$ and
\begin{equation}
I_{1}(z)=\frac{(2\pi)^{\frac{1-m}{2}}}{\rho
a^{\alpha}m^{\frac{1}{2}}}
G^{m+1,0}_{0,m+1}\biggl[\frac{ab^m}{m^m}\bigg|_{0,\frac{1}{m},\cdots,
\frac{m-1}{m},\alpha}\biggr],~~a>0,~b>0,~\alpha>0,~\rho>0.
\end{equation}

\noindent
The proof of the theorem will be given later.\\

\noindent
 Now we will consider a wider class of integrals as
extended forms of (1) and (2). Consider the integrals
\begin{equation}
I_{1\beta}^{(\delta)}=\int_{0}^{\infty}x^{\alpha-1}[1+a(\beta-1)x^\delta]^{-{\frac{1}
{\beta-1}}}{\rm e}^{-bx^{-\rho}}{\rm
d}x,~a>0,~b>0,~\delta>0,~\rho>0,~\beta>1,
\end{equation}
\begin{equation}
I_{2\beta}^{(\delta)}=\int_{0}^{d}x^{\alpha-1}[1-a(1-\beta)x^\delta]^{\frac{1}{1-\beta}}
{\rm e}^{-bx^{-\rho}}{\rm d}x,~a>0,~b>0,~\delta>0,~\rho>0,~\beta<1.
\end{equation}
\\
These are obtained by replacing $e^{-ax^{\delta}}$ by
$[1-a(1-\beta)x^\delta]^{\frac{1}{1-\beta}}$ so that when
$\beta\rightarrow1$ we have the integrals in (1)and (2). Thus
$I_{1\beta}^{(\delta)}$ and $I_{2\beta}^{(\delta)}$ are the extended
families of integrals in $I_1^{(\delta)}$ and $I_{2}^{(\delta)}$
respectively. Here we obtain the following main results for the
extended forms of the reaction
probability integrals. \\

\noindent
{\bf\Large{Theorem 2}}\\

\par A constant multiple of the extended reaction rate integral in
(7), under the condition $\frac{\delta}{\rho}=m,m=1,2,\cdots$, can
be obtained as a solution of
 the differential equation,
\begin{equation}
\bigl[{(-1)}^{m+1}z\bigl(\eta+\frac{1}{\beta-1}-\frac{\alpha}{m\rho}\bigr)-\eta\bigl
(\eta-\frac{1}{m}\bigr) \cdots\bigl(\eta-\frac{m-1}{m}\bigr)\bigl
(\eta-\frac{\alpha}{m\rho}\bigr)\bigr]f(z)=0,
\end{equation}
where,~$z=\frac{a(\beta-1)b^m}{m^m},~\eta=z\frac{{\rm d}}{{\rm
dz}},~f(z)=\frac{\rho[a(\beta-1)]^{\frac{\alpha}{m\rho}}\Gamma\bigl(\frac{1}{\beta-1}
\bigr)m^{\frac{1}{2}}} {(2\pi)^{\frac{1-m}{2}}}
I_{1\beta}^{m\rho}(z)$ and
\begin{equation}
I_{1\beta}^{m\rho}(z)=\frac{{(2\pi)}^{\frac{1-m}{2}}}{\rho{[a(\beta-1)]}^{\frac{\alpha
}{m\rho}} \Gamma\bigl(\frac{1}{\beta-1}\bigr)m^{\frac{1}{2}}}
G_{1,m+1}^{m+1,1}\biggl[\frac{a(\beta-1)b^m}{m^m}\bigg|_
{0,\frac{1}{m},\cdots,\frac{m-1}{m}, \frac{\alpha}{m
\rho}}^{\frac{\beta-2}{\beta-1}+\frac{\alpha}{m\rho}}\biggr],
\end{equation}
$~a>0,~b>0,~\alpha>0, ~\rho>0,~\beta>1$.\\

\noindent
{\bf\Large{Corollary 2}}\\
\par When $\delta=1$, a constant multiple of the extended reaction
rate integral in (7), under the condition
$\frac{1}{\rho}=m,m=1,2,\cdots,$ can be obtained as a solution of
the differential equation,
\begin{equation}
\bigl[{(-1)}^{m+1}z\bigl(\eta+\frac{1}{\beta-1}-\alpha\bigr)-\eta\bigl
(\eta-\frac{1}{m}\bigr) \cdots\bigl(\eta-\frac{m-1}{m}\bigr)\bigl
(\eta-\alpha\bigr)\bigr]f(z)=0,
\end{equation}
where,~$z=\frac{a(\beta-1)b^m}{m^m},~\eta=z\frac{{\rm d}}{{\rm dz}},
~f(z)=\frac{\rho[a(\beta-1)]^{\alpha}\Gamma\bigl(\frac{1}{\beta-1}\bigr)
m^{\frac{1}{2}}} {(2\pi)^{\frac{1-m}{2}}} I_{1\beta}(z)$ and
\begin{equation}
I_{1\beta}(z)=\frac{{(2\pi)}^{\frac{1-m}{2}}}{\rho{[a(\beta-1)]}^{\alpha}\Gamma\bigl
(\frac{1} {\beta-1}\bigr)m^\frac{1}{2}}
G_{1,m+1}^{m+1,1}\biggl[\frac{a(\beta-1)b^m}{m^m}\bigg|_{0,\frac{1}{m},
\cdots,\frac{m-1}{m},
\alpha}^{\frac{\beta-2}{\beta-1}+\alpha}\biggr],
\end{equation}
$~a>0,~b>0,~\alpha>0, ~\rho>0,~\beta>1$.\\

\noindent
{\bf\Large{Theorem 3}}\\
\par A constant multiple of the extended reaction rate integral in
(8), under the condition $\frac{\delta}{\rho}=m,m=1,2,\cdots$, can
be obtained as a solution of the differential equation,
\begin{equation}
\bigl[{(-1)}^{m}z\bigl(\eta-\frac{1}{1-\beta}-\frac{\alpha}{m\rho}\bigr)-\eta\bigl
(\eta-\frac{1}{m}\bigr)
\cdots\bigl(\eta-\frac{m-1}{m}\bigr)\bigl
(\eta-\frac{\alpha}{m\rho}\bigr)\bigr]f(z)=0,
\end{equation}
where,~$z=\frac{a(1-\beta)b^m}{m^m},~\eta=z\frac{{\rm d}}{{\rm
dz}},~
f(z)=\frac{\rho[a(1-\beta)]^{\frac{\alpha}{m\rho}}m^{\frac{1}{2}}}{\Gamma\bigl
(\frac{2-\beta}{1-\beta}\bigr)(2\pi)^{\frac{1-m}{2}}}
I_{2\beta}^{m\rho}(z)$ and
\begin{equation}
I_{2\beta}^{m\rho}(z)=\frac{\Gamma\bigl(\frac{2-\beta}{1-\beta}\bigr){(2\pi)}
^{\frac{1-m}{2}}}
{\rho{[a(1-\beta)]}^{\frac{\alpha}{m\rho}}m^{\frac{1}{2}}}
G_{1,m+1}^{m+1,0}\biggl[\frac{a(1-\beta)b^m}{m^m}\bigg|_{0,\frac{1}{m},
\cdots,\frac{m-1}{m}, \frac{\alpha}{m
\rho}}^{\frac{2-\beta}{1-\beta}+\frac{\alpha}{m\rho}}\biggr],
\end{equation}
$~a>0,~b>0,~\alpha>0, ~\rho>0,~\beta<1$.\\

\noindent
{\bf\Large{Corollary 3}}\\

\par When $\delta=1$ a constant multiple of the extended reaction rate
integral in (8), under the condition
$\frac{1}{\rho}=m,m=1,2,\cdots$, can be obtained as a solution of
the differential equation,
\begin{equation}
\bigl[{(-1)}^{m}z\bigl(\eta-\frac{1}{1-\beta}-\alpha\bigr)-\eta\bigl
(\eta-\frac{1}{m}\bigr)
\cdots\bigl(\eta-\frac{m-1}{m}\bigr)\bigl (\eta-\alpha\bigr)\bigr]f(z)=0,
\end{equation}
where,~$z=\frac{a(1-\beta)b^m}{m^m},~\eta=z\frac{{\rm d}}{{\rm
dz}},~f(z)=\frac{\rho[a(1-\beta)]^{\alpha}m^{\frac{1}{2}}}{\Gamma\bigl(\frac{2-\beta}
{1-\beta}\bigr)(2\pi)^{\frac{1-m}{2}}} I_{2\beta}(z)$ and
\begin{equation}
I_{2\beta}(z)=\frac{\Gamma\bigl(\frac{2-\beta}{1-\beta}\bigr){(2\pi)}^{\frac{1-m}{2}}}
{\rho{[a(1-\beta)]}^{\alpha}m^\frac{1}{2}}
G_{1,m+1}^{m+1,0}\biggl[\frac{a(1-\beta)b^m}{m^m}\bigg|_{0,\frac{1}{m},
\cdots,\frac{m-1}{m},
\alpha}^{\frac{2-\beta}{1-\beta}+\alpha}\biggr],
\end{equation}
$~a>0,~b>0,~\alpha>0, ~\rho>0,~\beta<1$.\\
\\
\noindent The proofs of the above theorems will be given after
evaluating the integrals first.
\section{Evaluation of the integral $I_{1}^{(\delta)}$ }
The following procedure is available in Mathai and Haubold (1988).
For the sake of completeness a brief outline of the derivation will
be given here. \noindent
\begin{equation}
I_{1}^{(\delta)}=\int_{0}^{\infty}x^{\alpha-1}\rm
{e}^{-ax^{\delta}-bx^{-\rho}}{\rm
d}x,~~a>0,~b>0,~\delta>0,~\rho>0.\nonumber
\end{equation}
 Here the integrand can be taken as a product of
positive integrable functions and then we can apply statistical distribution
theory to evaluate this integral. Let $x_1$ and $x_2$ be real scalar independent
random variables having densities
\begin{equation}
f_1(x_1)=\left\{\begin{array}{ll}
c_1x_1^{\alpha}{\rm e}^{-a{x_1^\delta}},& 0<x_1<\infty,~a>0,~\delta>0\\
0,& \text{elsewhere}\end{array}\right.
\end{equation}
and
\begin{equation}
f_2(x_2)=\left\{\begin{array}{ll}
c_2{\rm e}^{-x_{2}^{\rho}},& 0<x_2<\infty,~\rho>0\\
0,& \text{elsewhere,}\end{array}\right.
\end{equation}
where $c_1$ and $c_2$ are normalizing constants. Let us transform $x_1$ and $x_2$
to $u=x_1x_2$ and $v=x_1$. Then the marginal density of $u $ is given by
\begin{eqnarray}
g_1(u)&=&\int_{v}\frac{1}{v}f_1(v)f_2(\frac{u}{v}){\rm d}v\\&=&
c_1c_2\int_{0}^{\infty}v^{\alpha-1} \rm {e}^{-av^{\delta}-bv^{-\rho}}{\rm
d}v,~where~b=u^\rho,~\delta>0,~\rho>0.
\end{eqnarray}
Let us evaluate the density through expected values or moments.
\begin{equation}
E(u^{s-1})=E(x_1^{s-1})E(x_2^{s-1})
\end{equation}
due to statistical independence of $x_1$ and $x_2$.
\begin{equation*}
E(x_1^{s-1})=c_1\int_{0}^{\infty}x_1^{\alpha+s-1}{\rm
e}^{-ax_1^\delta}{\rm d}x_1.
\end{equation*}
Putting $y=ax_1^\delta$ and evaluating the integral as a gamma integral, one has,
\begin{equation}
E(x_1^{s-1})=\frac{c_1}{\delta
a^{\frac{\alpha}{\delta}+\frac{s}{\delta}}}\Gamma
(\frac{\alpha}{\delta}+\frac{s}{\delta}),~~\Re({\alpha+s})>0,
\end{equation}
where $\Re(.)$ denotes the real part of (.).
\begin{equation*}
E(x_2^{s-1})=c_2\int_{0}^{\infty}x_2^{s-1}{\rm e}^{-x_2^\rho}{\rm
d}x_2.
\end{equation*}
Putting $y=x_2^\rho$, we get
\begin{equation}
E(x_2^{s-1})=\frac{c_2}{\rho}\Gamma(\frac{s}{\rho}),~~\Re(s)>0.
\end{equation}
From (22) and (23)
\begin{equation}
E(u^{s-1})=\frac{c_1c_2}{\delta\rho
a^{\frac{\alpha}{\delta}+\frac{s}{\delta}}}\Gamma(\frac{s}{\rho})\Gamma
({\frac{\alpha}{\delta}+\frac{s}{\delta}}),
~\Re(s)>0,~\Re({\alpha+s})>0.
\end{equation}
Looking at the $(s-1)^{th}$ moment as the Mellin transform of the
corresponding density and then taking the inverse Mellin transform
we get the density of $u$,
\begin{equation}
g_1(u)=\frac{c_1c_2}{\delta\rho a^{\frac{\alpha}{\delta}}}\frac{1}{2\pi
i}\int_{c-i \infty}^{c+i
\infty}\Gamma(\frac{s}{\rho})\Gamma(\frac{\alpha}{\delta}+\frac{s}{\delta})
(a^{\frac{1}{\delta
}}b^{\frac{1}{\rho}})^{-s}{\rm d}s.
\end{equation}
Comparing (20) and (25)
\begin{eqnarray}
I_{1}^{(\delta)}&=&\int_{0}^{\infty}x^{\alpha-1}e^{-ax^{\delta}-bx^{-\rho}}{\rm
d}x\nonumber\\
&=&\frac{1}{\delta\rho a^{\frac{\alpha}{\delta}}}\frac{1}{2\pi i}\int_{c-i
\infty}^{c+i
\infty}\Gamma(\frac{s}{\rho})\Gamma(\frac{\alpha}{\delta}+\frac{s}{\delta})
(a^{\frac{1}{\delta
}}b^{\frac{1}{\rho}})^{-s}{\rm d}s.
\end{eqnarray}
This contour integral can be written as an $H$-function (Mathai and
Saxena, 1978). That is,
\begin{equation}
I_{1}^{(\delta)}=\frac{1}{\delta\rho
a^{\frac{\alpha}{\delta}}}H_{0,2}^{2,0}\biggl[a^{\frac{1}{\delta}}b^{\frac{1}
{\rho}}\bigg|_{(0,\frac{1}{\rho}),(\frac{\alpha}{\delta},
\frac{1}{\delta})}\biggr],~~a>0,~b>0,~\alpha>0,~\delta>0,~\rho>0.
\end{equation}
Make the transformation $\frac{s}{\delta}=s_1$ in (26)
\begin{equation}
I_{1}^{(\delta)}=\frac{1}{\rho a^{\frac{\alpha}{\delta}}}\frac{1}{2\pi i}\int_{c-i
\infty}^{c+i \infty}\Gamma(\frac{\delta
s_1}{\rho})\Gamma(\frac{\alpha}{\delta}+s_1)(ab^{\frac{\delta}{\rho}})^{-s_1}{\rm
d}s_1.
\end{equation}
Let us consider the special case where $\frac{\delta}{\rho}=m,~
m=1,2,\cdots$. In physics problems $\rho=\frac{1}{2}$ and $\delta$
an integer. Then this assumption of $\frac{\delta}{\rho}=m,~
m=1,2,\cdots$ is meaningful at least in some physical problems.
That is,
\begin{equation*}
I_{1}^{(m\rho)}=\frac{1}{\rho a^{\frac{\alpha}{m\rho}}}\frac{1}{2\pi i}
\int_{c_1-i \infty}^{c_1+i \infty}\Gamma( ms_1)\Gamma
\bigl(\frac{\alpha}{m\rho}+s_1\bigr)(ab^m)^{-s_1}{\rm d}s_1.
\end{equation*}
This can be reduced to a $G$-function by using the multiplication formula for
gamma functions, namely,
\begin{equation}
\Gamma(mz)=(2\pi)^{\frac{1-m}{2}}m^{mz-\frac{1}{2}}\Gamma(z)\Gamma
\bigl(z+\frac{1}{m} \bigr )\cdots\Gamma\bigl(z+\frac{m-1}{m}\bigr)
,~m=1,2,\cdots~.
\end{equation}
Then we have,
\begin{equation}
I_{1}^{(m\rho)}=\frac{(2\pi)^{\frac{1-m}{2}}}{\rho
a^{\frac{\alpha}{m \rho}}m^{\frac{1}{2}}} \frac{1}{2\pi i}
\int_{c_1-i \infty}^{c_1+i \infty}m^{ms_1}\Gamma( s_1)\Gamma
\bigl(s_1+\frac{1}{m}\bigr)\cdots\Gamma
\bigl(s_1+\frac{m-1}{m}\bigr)\Gamma\bigl(s_1+
\frac{\alpha}{m\rho}\bigr)(ab^m)^{-s_1}{\rm d}s_1.
\end{equation}\\
On evaluating (30), we get (4). (see
Mathai and Haubold, 1988; Saxena, 1960); Haubold and Mathai, 1984).\\

\noindent {\bf{Particular case, $\delta=1$}}\\

\noindent When $\delta=1$, we get the reaction rate integral in (1),
under the condition
 $\frac{1}{\rho}=m,$\\
$ m=1,2,\cdots$ as,
\begin{equation}
I_1=\int_{0}^{\infty}x^{\alpha-1}e^{-ax-bx^{-\rho}}{\rm
d}x,~~a>0,~b>0, ~\rho>0.
\end{equation}\\
On evaluating this integral, we get (6). Observe that $m=2$ is a
real physical situation, see for example, Mathai and Haubold (1988).

\section{Evaluation of extended integrals}
We have,
\begin{equation*}
I_{1\beta}^{(\delta)}=\int_{0}^{\infty}x^{\alpha-1}[1+a(\beta-1)x^\delta]^{-{\frac{1}
{\beta-1}}}{\rm e}^{-bx^{-\rho}}{\rm
d}x,~a>0,~b>0,~\delta>0,~\rho>0,~\beta>1.
\end{equation*}
As $\beta \rightarrow 1$,
$[1-a(1-\beta)x^\delta]^{\frac{1}{1-\beta}}$ becomes
$e^{-ax^{\delta}}$ so that we can extend the reaction rate integrals
in (1) and (2) using the pathway parameter $\beta$.
Then here arise two cases:(i) $\beta<1$,(ii) $\beta>1$.\\

\noindent
{\bf {Case(i), $\beta<1$}}\\
\begin{equation*}
I_{2\beta}^{(\delta)}=\int_{0}^{d}x^{\alpha-1}[1-a(1-\beta)x^\delta]^
{\frac{1}{1-\beta}}{\rm e}^{-bx^{-\rho}}{\rm
d}x,~a>0,~b>0,~\delta>0,~\beta<1.
\end{equation*}
Here let us take
\begin{equation}
f_1(x_1)=\left\{\begin{array}{ll}
c_1x_1^{\alpha}[1-a(1-\beta){x_1}^\delta]^{\frac{1}{1-\beta}},
& 0<x_1<\bigl[\frac{1}{a(1-\beta)}\bigr]^{\frac{1}
{\delta}},~a>0,~\delta>0,~\beta<1\\
0,& \text{elsewhere}\end{array}\right.
\end{equation}
and
\begin{equation}
f_2(x_2)=\left\{\begin{array}{ll}
c_2{\rm e}^{-x_{2}^{\rho}},& 0<x_2<\infty,~\rho>0\\
0,& \text{elsewhere,}\end{array}\right.
\end{equation}
where $c_1$ and $c_2$ are the normalizing constants. Let us consider
the case where
$d=\bigl[\frac{1}{a(1-\beta)}\bigr]^{\frac{1}{\delta}}$
in (8). \\
Note that $f_1(x_1)$ in (32) with $\alpha=0$ and $\delta=1$ is
Tsallis statistics leading to nonextensive statistical mechanics.
Also observe that the functional part of (32) for $\alpha=0$
and $\delta=1$ gives the power law as well.
\begin{equation}
{\rm \frac{d}{dx}}\bigl[ \frac{f_1(x_1)}{c_1}\bigr]=-{\bigl[
\frac{f_1(x_1)}{c_1}\bigr]}^\beta.
\end{equation}\\
Note that $f_2(x_2)$ in (33) is what is known as stretched
exponential in physics literature. Thus the extended reaction rate
integral in (8) is the Mellin convolution of Tsallis nonextensive
statistics and stretched exponentials. The starting publication of
nonextensive statistical mechanics may be seen from Tsallis (1988).
Then proceeding as before, we get,
\begin{eqnarray}
I_{2\beta}^{(\delta)}&=&\int_{0}^{\bigl[\frac{1}{a(1-\beta)}\bigr]^{\frac{1}
{\delta}}}x^{\alpha-1}[1-a(1-\beta)x^\delta]^
{\frac{1}{1-\beta}}{\rm e}^{-bx^{-\rho}}{\rm d}x\\
&=&\frac{\Gamma\big(1+\frac{1}{1-\beta}\bigr)}{\rho\delta{[a(1-\beta)]
}^{\frac{\alpha}{\delta}}} \frac{1}{2\pi i}\int_{c-i \infty}^{c+i
\infty}\frac{\Gamma(\frac{s}{\rho})\Gamma(\frac{\alpha}{\delta}+
\frac{s}{\delta})[a^{\frac{1}{\delta
}}{(1-\beta)}^{\frac{1}{\delta}}b^{\frac{1}{\rho}}]^{-s}{\rm
d}s}{\Gamma\bigl(\frac{\alpha}{\delta}+1+\frac{1}{1-\beta}+\frac{s}{\delta}\bigr)}\\
&=&\frac{\Gamma\big(1+\frac{1}{1-\beta}\bigr)}{\rho\delta{\bigl[a(1-\beta)\bigr]}^
{\frac{\alpha}{\delta}}}
H_{1,2}^{2,0}\biggl[a^{\frac{1}{\delta}}{(1-\beta)}^{\frac{1}{\delta}}b^{\frac{1}{\rho}}
\bigg|_{(0,\frac{1}{\rho}) ,(\frac{\alpha}{\delta},
\frac{1}{\delta})}^{(\frac{\alpha}{\delta}+1+\frac{1}{1-\beta},\frac{1}{\delta})}
\biggr],
\end{eqnarray}
$a>0,~b>0,~\alpha>0,\delta>0~\rho>0,~\beta<1,~\Re{(s)}>0,~\Re({\alpha+s})>0$.\\
Make the transformation $\frac{s}{\delta}=s_1$ in (36)
\begin{equation}
I_{2\beta}^{(\delta)}=\frac{\Gamma\bigl(1+\frac{1}{1-\beta}\bigr)}
{\rho{\bigl[a(1-\beta)\bigr]}^{\frac{\alpha}{\delta}}} \frac{1}{2\pi
i}\int_{c-i \infty}^{c+i \infty}\frac{\Gamma(\frac{\delta
s_1}{\rho})\Gamma(\frac{\alpha}{\delta}+s_1)[a
(1-\beta)b^{\frac{\delta}{\rho}}]^{-s_1}{\rm
d}s_1}{\Gamma\bigl(\frac{\alpha}{\delta}+1+\frac{1}{1-\beta}+s_1
\bigr)}.
\end{equation}
Let us consider the case where $\frac{\delta}{\rho}=m,~
m=1,2,\cdots$. Then we get (14).\\

\noindent {\bf{Lemma 1}}: As $\beta \rightarrow 1$,
$I_{2\beta}^{(\delta)}$ becomes
$I_2^{(\delta)}$.\\

\noindent
{\bf{Proof}}:\\
\begin{equation}
\displaystyle \lim_{\beta\rightarrow 1}I_{2\beta}^{(\delta)}=
\displaystyle \lim_{\beta\rightarrow 1}\frac{1}{2 \pi i}\int_{c-i
\infty}^{c+i
\infty}\frac{\Gamma(\frac{s}{\rho})\Gamma(\frac{\alpha}{\delta}+
\frac{s}{\delta})(a^{\frac{1}{\delta
}}b^{\frac{1}{\rho}})^{-s}}{\rho \delta a^{\frac{\alpha}{\delta}}}
\frac{\Gamma\bigl(1+\frac{1}{1-\beta}\bigr)(1-\beta)^{-(\frac{\alpha}{\delta}+\frac{s}
{\delta})}}{\Gamma
\bigl(\frac{\alpha}{\delta}+\frac{s}{\delta}+1+\frac{1}{1-\beta}
\bigr)}{\rm d}s.
\end{equation}
Now apply the asymptotic formula for gamma functions, namely, for
$|z|\rightarrow \infty$ and $a$ is bounded (Mathai, 1993),
\begin{equation}
\Gamma(z+a)\rightarrow(2\pi)^{\frac{1}{2}}z^{z+a-\frac{1}{2}}{\rm
e}^{-z}.
\end{equation}
 Apply this to the gamma ratios in (39) by taking $z$ as
$\frac{1}{1-\beta}$ and $a$ as 1 and
$\frac{\alpha}{\delta}+\frac{s}{\delta}+1$ respectively. Then,
\begin{equation}
\displaystyle \lim_{\beta\rightarrow
1}\frac{\Gamma\bigl(1+\frac{1}{1-\beta}\bigr)(1-\beta)^{-(\frac{\alpha}{\delta}+\frac{s}
{\delta})}}{\Gamma
\bigl(\frac{\alpha}{\delta}+\frac{s}{\delta}+1+\frac{1}{1-\beta}
\bigr)}=1
\end{equation}
Hence\\
\begin{equation}
\displaystyle \lim_{\beta\rightarrow
1}I_{2\beta}^{(\delta)}=I_2^\delta
\end{equation}
which establishes the result.\\

\noindent
{\bf{Case(ii), $\beta>1$}}\\
\begin{equation*}
I_{1\beta}^{(\delta)}=\int_{0}^{\infty}x^{\alpha-1}[1+a(\beta-1)x^\delta]^
{-{\frac{1}{\beta-1}}}{\rm e}^{-bx^{-\rho}}{\rm
d}x,~a>0,~b>0,~\delta>0,~\rho>0,~\beta>1.
\end{equation*}
Here let us take
\begin{equation}
f_1(x_1)=\left\{\begin{array}{ll}
c_1x_1^{\alpha}[1+a(\beta-1){x_1}^\delta]^{-{\frac{1}{\beta-1}}},
& 0<x_1<\infty,~a>0,~\delta>0,~\beta>1\\
0,& \text{elsewhere}\end{array}\right.
\end{equation}
and
\begin{equation}
f_2(x_2)=\left\{\begin{array}{ll}
c_2{\rm e}^{-x_{2}^{\rho}},& 0<x_2<\infty,~\rho>0\\
0,& \text{elsewhere,}\end{array}\right.
\end{equation}
where $c_1$ and $c_2$ are the normalizing constants. Observe that
$f_1(x_1)$ of (43) is nothing but the superstatistics of Beck and
Cohen (2003), and the density in (44) is the stretched exponential.
Hence (7) can be looked upon as the Mellin convolution of
superstatistics and stretched exponentials. A large number of
published articles are there on superstatistics. Then proceeding as
before, we get,
\begin{eqnarray}
I_{1\beta}^{(\delta)}&=&\int_{0}^{\infty}x^{\alpha-1}[1+a(\beta-1)x^\delta]
^{-{\frac{1}{\beta-1}}}{\rm
e}^{-bx^{-\rho}}{\rm d}x\nonumber\\
&=&K~H_{1,2}^{2,1}\biggl[a^{\frac{1}{\delta}}{(\beta-1)}^{\frac{1}{\delta}}b^
{\frac{1}{\rho}}\bigg|_{(0,\frac{1}{\rho}) ,(\frac{\alpha}{\delta},
\frac{1}{\delta})}^{(\frac{\alpha}{\delta}+\frac{\beta-2}{\beta-1},\frac{1}
{\delta})}\biggr],
\end{eqnarray}
\noindent where
$K=\frac{1}{\Gamma\bigl(\frac{1}{\beta-1}\bigr)\rho\delta{[a(\beta-1)]}^
{\frac{\alpha}{\delta}}},
~a>0,~b>0,~\alpha>0,~\rho>0,~\delta>0,~\beta>1,~\Re{(s)}>0,\\~\Re({\alpha+s})>0.$
\noindent Make the transformation $\frac{s}{\delta}=s_1$ and let
$\frac{\delta}{\rho}=m,~m=1,2,\cdots$ in (45). Then we get (10).\\

\noindent {\bf{Lemma 2}}: As $\beta \rightarrow 1$,
$I_{1\beta}^{(\delta)}$ becomes $I_1^{(\delta)}$.\\
\newpage
\noindent
{\bf{Particular case, $\delta=1$}}\\

\noindent When $\delta=1$, we get the reaction rate integral under
the condition
$\frac{1}{\rho}=m,$\\
$ m=1,2,\cdots$ as,
\begin{equation}
I_{2\beta}=\int_{0}^{\bigl[\frac{1}{a(1-\beta)}\bigr]}x^{\alpha-1}[1-a(1-\beta)x]^
{\frac{1}{1-\beta}}{\rm e}^{-bx^{-\rho}}{\rm d}x,~~a>0,~b>0,
~\rho>0,~\beta<1,
\end{equation}\\
on evaluating this integral we get (16),
and\\
\begin{equation}
I_{1\beta}=\int_{0}^{\infty}x^{\alpha-1}[1+a(\beta-1)x]^{\frac{-1}{\beta-1}}{\rm
e}^{-bx^{-\rho}}{\rm d}x,~~a>0,~b>0, ~\rho>0,~\beta>1,
\end{equation}\\
on evaluating this integral we get (12).\\

 \noindent
{\bf\large{Proof of Theorem 1}}\\

\par For proving the theorem we will make use of the fact that the reaction
probability integral as well as the extended reaction probability
integrals, as given in (1),(2),(7),(8), under the condition
$\delta=m\rho,~m=1,2,\cdots$ can be written in terms of
$G$-functions. Hence when this condition is satisfied we can invoke
the properties of $G$-functions. \noindent It is well known that the
$G$-function defined
by\\
\begin{eqnarray}
G_{p,q}^{m,n}\bigl[z\big|_{b_1,,...,b_p}^ {a_1,...,a_p}\bigr]&=&\frac{1}{2\pi
i}\int_L\phi (s)~z^{-s}{\rm d}s,
\end{eqnarray}
where
\begin{eqnarray}
 \phi (s)&=&\frac{\bigl\{\prod _{j=1}^m \Gamma
(b_j+ s)\bigr\}~\bigl\{\prod _{j=1}^n\Gamma (1-a_j- s)\bigr\}}
 {\bigl\{\prod _{j=m+1}^q\Gamma (1-b_j- s)\bigr\}~
\bigl\{ \prod _{j=n+1}^p \Gamma (a_j+s)\bigr\}}\nonumber,
\end{eqnarray}
$a_j,~j=1,2,...,p$ and $b_j,~j=1,2,...,q$ are complex numbers, $L$ is a contour
separating the poles of $\Gamma (b_j+s),~j=1,2,...,m$ from those of $\Gamma
(1-a_j-s),~j=1,2,...,n$,\\
satisfies the following differential equation.
\begin{equation}
\bigl[{(-1)}^{p-m-n}z\prod_{j=1}^{p}(\eta-a_j+1)-\prod_{j=1}^{q}(\eta-b_j)\bigr]G(z)=0,
~\eta=z\frac{{\rm
d}}{{\rm dz}}.
\end{equation}
\newpage
\noindent
 This equation is intuitively evident from the following
facts:
\begin{equation}
\eta z^{-s}=z\frac{{\rm d}}{{\rm dz}}(z^{-s})=(-s)z^{-s};\nonumber
\end{equation}
\begin{equation}
(\eta-a_j+1)z^{-s}=(1-a_j-s)z^{-s};\nonumber
\end{equation}
\begin{equation}
(1-a_j-s)\Gamma(1-a_j-s)=\Gamma(2-a_j-s),\nonumber
\end{equation}
(see Mathai, 1993). The depleted case of the reaction rate integral
\begin{equation}
I_{1}^{(\delta)}=\int_{0}^{\infty}x^{\alpha-1}e^{-ax^{\delta}-bx^{-\rho}}{\rm
d}x,~~a>0,~b>0,~\delta>0,~\rho>0
\end{equation} can be expressed in terms of $G$-functions under the conditions
$\frac{\delta}{\rho}=m,m=1,2,\cdots$  shown in (4). That is,
\begin{equation}
{(2\pi)}^{\frac{m-1}{2}}m^{\frac{1}{2}}\rho
a^{\frac{\alpha}{m\rho}}I_{1}^{(m\rho)}=
G^{m+1,0}_{0,m+1}\biggl[\frac{ab^m}{m^m}\bigg|_{0,\frac{1}{m},\cdots,\frac{m-1}{m},
\frac{\alpha}{m \rho}}\biggr],
\end{equation}\\
$~~~~~~~~~~~~~~~~~~~~~~~~~~~~~~~~~~~~~~~~~~~~~~~~~~a>0,~b>0,~\alpha>0,~\delta>0,
~\rho>0,~\frac{\delta}{\rho}=m, ~m=1,2,\cdots$.\\The $G$-function in
(51), satisfies the differential equation in (3). So the left hand
side of (51) also satisfies the differential equation (3). Similar
are the proofs of the other theorems and hence deleted.\\

\noindent
{\bf{\large Acknowledgment}}\\

 The authors would like to thank the Department of Science and Technology,
Government of India, New Delhi, for the financial assistance for this work under
project-number SR/S4/MS:287/05 and the Centre for
Mathematical Sciences for providing all facilities.\\

\noindent {\bf{\large References}}
\begin{enumerate}[{[1]}]
\item
Beck, C. and Cohen, E.G.D.: Superstatistics. Physica A {\bf 322} 267-275 (2003)
\item
Haubold, H.J. and Mathai, A.M.: On nuclear reaction rate theory. Annalen der Physik (Berlin) {\bf 41} 380-396 (1984)
\item
Haubold, H.J. and Mathai, A.M.: On thermonuclear reaction rates. Astrophysics and Space Science {\bf 258} 185-199 (1998)
\item
Luke, Y.L.: The Special Functions and Their Approximations. Academic Press, New York (1969)
\item
Mathai, A.M., Saxena, R.K.: The $H$-function with Applications in Statistics and Other Disciplines. Wiley Halsted, New York, London and Sidney (1978)
\item
Mathai, A.M.: A Handbook of Generalized Special Functions for Statistical and Physical Sciences. Oxford University Press, Oxford (1993)
\item
Mathai, A.M. and Haubold, H.J.: Modern Problems in Nuclear and Neutrino Astrophysics. Akademie-Verlag Berlin (1988)
\item
Saxena, R.K.: Some theorems on generalized Laplace transform-1. Proceedings of the National Institute of Science India {\bf A 26} 400-413 (1960)
\item
Tsallis, C.: Possible generalizations of Boltzmann-Gibbs statistics. Journal of Statistical Physics {\bf 52} 479-487 (1988)

\end{enumerate}
\end{document}